\documentclass[11pt,superscriptaddress,aps,prd,preprint]{revtex4}
\everymath{\displaystyle}
\usepackage{graphicx}
\usepackage[T1]{fontenc}
\usepackage{amsmath}
\usepackage{amssymb}
\usepackage{graphicx}
\newcommand{\bea}{\begin{eqnarray}}
\newcommand{\eea}{\end{eqnarray}}
\begin{document}

\title{On Regular Black Holes at Finite Temperature}

\author{S. C. Ulhoa}
\email{sc.ulhoa@gmail.com}
\affiliation{International Center of Physics, Instituto de F\'isica, Universidade de Bras\'ilia, 70910-900, Bras\'ilia, DF, Brazil}

\author{E. P. Spaniol}
\email{spaniol.ep@gmail.com} \affiliation{UDF Centro
Universit\'ario and Centro Universit\'ario de Bras\'ilia UniCEUB, Bras\'ilia, DF, Brazil.}

\author{R. Gomes}
\email{rogosomuza@gmail.com}
\affiliation{Instituto de F\'isica, Universidade de Bras\'ilia, 70910-900, Bras\'ilia, DF, Brazil}

\author{A. F. Santos}
\email{alesandroferreira@fisica.ufmt.br}
\affiliation{Instituto de F\'{\i}sica, Universidade Federal de Mato Grosso,\\
78060-900, Cuiab\'{a}, Mato Grosso, Brazil}

\author{A. E. Santana}
\email{a.berti.santana@gmail.com}
\affiliation{International Center of Physics, Instituto de F\'isica, Universidade de Bras\'ilia, 70910-900, Bras\'ilia, DF, Brazil}

\begin{abstract}

The Thermo Field Dynamics (TFD) formalism is used to investigate the regular black holes at finite temperature. Using the Teleparalelism Equivalent to General Relativity (TEGR) the gravitational Stefan-Boltzmann law and the gravitational Casimir effect at zero and finite temperature are calculated. In addition, the first law of thermodynamics is considered. Then the gravitational entropy and the temperature of the event horizon of a class of regular black holes are determined.

\end{abstract}
\maketitle

\section{Introduction} \label{sec.1}

The existence of singularities in the theory of general relativity has been problematic since its inception, especially those linked to black holes. The so-called fundamental singularities do not allow the application of the laws of physics, which are a type of difficult giving rise to some proposals to address the problem. One of them is the well-known cosmic censorship that was proposed by Penrose in the last century \cite{penrose1999question}. Another more particular proposal came with the solution of regular black holes whose first solution was obtained by Bardeen \cite{bardeen}. It is interesting to note that solutions describing regular black holes have an event horizon and therefore share many features with singular black holes. So if black holes are real objects, there is a good chance that the regular ones are the objects that will be experimentally perceived. In this way the theoretical study of such solutions becomes very relevant. Particularly the analysis of the thermodynamics of regular black holes can reveal substantial experimental implications. For this it is necessary to define how the temperature is introduced and how from it the gravitational entropy is obtained. In addition, other effects can be predicted by means of this termalization, such as the gravitational Stefan-Boltzmann law and Casimir effect. In this sense we will work with Thermo Field Dynamics (TFD) because it has proved to be a very powerful tool to deal with the termalization of a given field \cite{tfd,tfd2}. This approach requires doubling the Fock space which allows both time and temperature to be system variables. This is an advantage over Matsubara's approach \cite{mat}. Thermo Field Dynamics also requires the characterization of the field under analysis by means of the corresponding Green function. Regarding the gravitational field we will use an alternative description that is known as Teleparalelism Equivalent to General Relativity (TEGR). When the gravitational field is described in this alternative way some unique predictions are revealed.

Teleparalelism Equivalent to General Relativity is described in terms of tetrads which has many advantages over the metric formulation of gravitation as General Relativity, since equivalence occurs only in relation to field equations. Among them the most notorious is the existence of a gravitational energy that is obtained naturally through the Hamiltonian formulation of the TEGR \cite{maluf}. Moreover, the propagator of graviton obtained in general relativity does not coincide with that predicted by the TEGR \cite{usk}. This opens up scope for exploration of regular black holes via Thermo Field Dynamics. In this paper we will use TFD to calculate the Stefan-Boltzmann law and the Casimir effect through the TEGR.

This article is organized as follows. In section \ref{sec.3} Thermo Field Dynamics is introduced. In section \ref{sec.2} the ideas of the  Teleparallelism Equivalent to General Relativity are briefly recalled. In section \ref{sec.4} the gravitational Stefan-Boltzmann and entropy are calculated for regular black holes. In addition we also calculate for the same gravitational system the Casimir effect which is given at zero and finite temperature. Finally in the last section we present our conclusions. In this article we use the natural unities system, $G=c=1$. We denote the Lorentz symmetry by Latin indices, $a=(0),(1),(2),(3)$, while diffeomorphisms are denoted by greek indices, $\mu=0,1,2,3$.

\section{Thermo Field Dynamics (TFD)} \label{sec.3}

TFD is a thermal quantum field theory \cite{Umezawa1,tfd, Umezawa2, Umezawa22, Khanna1, Khanna2, Santana1, Santana2}. This formalism is used when it is desirable to have explicit time dependence in addition to the temperature. TFD is based on two basic ingredients: (i) a doubling of the Hilbert space, ${\cal S}$, of the original field system, giving rise to ${\cal S}_T={\cal S}\otimes \tilde{\cal S}$, where $\tilde{\cal S}$ is the tilde (dual) space. This doubling is defined by the tilde conjugation rules. (ii) The Bogoliubov transformation which introduces thermal effects through a rotation between tilde ($\tilde{\cal S}$) and non-tilde (${\cal S}$) operators. These ingredients allow to interpret the statistical average of an arbitrary operator $A$, as the expectation value in a thermal vacuum. The thermal vacuum, $|0(\beta) \rangle$, describes the thermal equilibrium of the system, where $\beta=\frac{1}{k_BT}$, $T$ is the temperature and $k_B$ is the Boltzmann constant.

By taking an arbitrary operator ${\cal A}$ and $\tilde{\cal A}$ in Hilbert space ${\cal S}$ and in tilde space $\tilde{\cal S}$, respectively, the Bogoliubov transformation is
\bea
\left( \begin{array}{cc} {\cal A}(\alpha)  \\\xi \tilde {\cal A}^\dagger(\alpha) \end{array} \right)={\cal U}(\alpha)\left( \begin{array}{cc} {\cal A}(k)  \\ \xi\tilde {\cal A}^\dagger(k) \end{array} \right),
\eea
where $\xi = -1$ for bosons and $\xi = +1$ for fermions. The Bogoliubov transformation, ${\cal U}(\alpha)$, is defined as
\bea
{\cal U}(\alpha)=\left( \begin{array}{cc} u(\alpha) & -w(\alpha) \\
\xi w(\alpha) & u(\alpha) \end{array} \right),
\eea
with $u^2(\alpha)+\xi w^2(\alpha)=1$. Here the $\alpha$ parameter is the compactification parameter defined by $\alpha=(\alpha_0,\alpha_1,\cdots\alpha_{D-1})$. The temperature effect is described by the choice $\alpha_0\equiv\beta$ and $\alpha_1,\cdots\alpha_{D-1}=0$, where $\beta=1/k_B T$ with $k_B$ being the Boltzmann constant.

Any field in the TFD formalism can be written in terms of the $\alpha $-parameter. As an example, consider the scalar field. Using the Bogoliubov transformation the scalar field  dependent of $\alpha$-parameter becomes
\bea
\phi(x;\alpha)&=&{\cal U}(\alpha)\phi(x){\cal U}^{-1}(\alpha),\nonumber\\
\tilde{\phi}(x;\alpha)&=&{\cal U}(\alpha)\tilde{\phi}(x){\cal U}^{-1}(\alpha).
\eea
Then the propagator for the scalar field in terms of $\alpha$-parameter is written as
\bea
G_0^{(AB)}(x-x';\alpha)&=&i\langle 0,\tilde{0}| \tau[\phi^A(x;\alpha)\phi^B(x';\alpha)]| 0,\tilde{0}\rangle\nonumber\\
&=&i\int \frac{d^4k}{(2\pi)^4}e^{-ik(x-x')}G_0^{(AB)}(k;\alpha),
\eea
where $A$ and $B$ represent the duplicate notation with $A \,\rm{and}\, B=1,2$ and $\tau$ is the time ordering operator. Here
\bea
G_0^{(AB)}(k;\alpha)={\cal U}^{-1}(\alpha)G_0^{(AB)}(k){\cal U}(\alpha).\label{phi}
\eea
It is important to note that, the physical quantities are given by the non-tilde variables. Using the Bogoliubov transformation in Eq. (\ref{phi}) the Green function becomes
\bea
G_0^{(11)}(k;\alpha)=G_0(k;\alpha)=G_0(k)+v^2(k;\alpha)[G_0(k)-G^*_0(k)],
\eea
with
\bea
G_0(k)=\frac{1}{k^2-m^2+i\epsilon}
\eea
and $v^2(k;\alpha)$ being the generalized Bogoliubov transformation \cite{GBT} which is given as
\bea
v^2(k;\alpha)=\sum_{s=1}^d\sum_{\lbrace\sigma_s\rbrace}2^{s-1}\sum_{l_{\sigma_1},...,l_{\sigma_s}=1}^\infty(-\eta)^{s+\sum_{r=1}^sl_{\sigma_r}}\,\exp\left[{-\sum_{j=1}^s\alpha_{\sigma_j} l_{\sigma_j} k^{\sigma_j}}\right],\label{BT}
\eea
where $d$ is the number of compactified dimensions, $\eta=1(-1)$ for fermions (bosons), $\lbrace\sigma_s\rbrace$ denotes the set of all combinations with $s$ elements and $k$ is the 4-momentum.

\section{Teleparalellism Equivalent to General Relativity (TEGR)}\label{sec.2}

General Relativity which is the standard approach for gravitation is based on Riemann geometry in which the field variables are the components of the metric tensor. Such a formulation doesn't lead to gravitational conserved quantities, partly because of the inclusion of local Lorentz symmetry and partly due to difficulty to formally establish a reference frame. Those problems are solved by TEGR which is formulated in terms of the tetrad field in a Weitzenb\"ock manifold. In the 1930 decade, Einstein tried to establish a unified field theory using the concept of distant teleparallelism \cite{einstein} which led to the formulation of a New General Relativity by Hyashi and Shirafuji \cite{hyashi}. Then a Hamiltonian formulation was successfully established which yielded conserved quantities such as the gravitational energy-momentum tensor and angular momentum \cite{maluf}.

Let us consider a Weitzenb\"ock manifold endowed with the Cartan connection \cite{cartan}
$$\Gamma_{\mu\lambda\nu}=e^{a}\,_{\mu}\partial_{\lambda}e_{a\nu},$$
then the associated torsion tensor is
\begin{equation}
T^{a}\,_{\lambda\nu}=\partial_{\lambda} e^{a}\,_{\nu}-\partial_{\nu}
e^{a}\,_{\lambda}\,. \label{4}
\end{equation}
Cartan connection is curvature free. On the other hand it is identically related to Christoffel symbols ${}^0\Gamma_{\mu \lambda\nu}$, which there exist in the realm of Riemannian geometry, by
\begin{equation}
\Gamma_{\mu \lambda\nu}= {}^0\Gamma_{\mu \lambda\nu}+ K_{\mu
\lambda\nu}\,, \label{2}
\end{equation}
where $K_{\mu \lambda\nu}$ is the contortion tensor which is defined by
\begin{eqnarray}
K_{\mu\lambda\nu}&=&\frac{1}{2}(T_{\lambda\mu\nu}+T_{\nu\lambda\mu}+T_{\mu\lambda\nu})\,,\label{3}
\end{eqnarray}
with $T_{\mu \lambda\nu}=e_{a\mu}T^{a}\,_{\lambda\nu}$.

In order to establish a Lagrangian density for TEGR we firstly note that the scalar curvature constructed out of the Christoffel symbols is written in terms of the torsion tensor, due to the identity in Eq. (\ref{3}), as
\begin{equation}
eR(e)\equiv -e({1\over 4}T^{abc}T_{abc}+{1\over
2}T^{abc}T_{bac}-T^aT_a)+2\partial_\mu(eT^\mu)\,.\label{eq5}
\end{equation}
Then getting rid of the total divergency which does not alters the field equations, we have
\begin{eqnarray}
\mathfrak{L}(e_{a\mu})&=& -\kappa\,e \Sigma^{abc}T_{abc} -\mathfrak{L}_M\;, \label{6}
\end{eqnarray}
where $\kappa=1/(16 \pi)$, $\mathfrak{L}_M$ is the Lagrangian
density of matter fields and $\Sigma^{abc}$ is defined by

\begin{equation}
\Sigma^{abc}={1\over 4} (T^{abc}+T^{bac}-T^{cab}) +{1\over 2}(
\eta^{ac}T^b-\eta^{ab}T^c)\;, \label{7}
\end{equation}
with $T^a=e^a\,_\mu T^\mu$. If a derivative with respect to tetrad field is performed in the Lagrangian density, it yields
\begin{equation}
\partial_\nu\left(e\Sigma^{a\lambda\nu}\right)={1\over {4\kappa}}
e\, e^a\,_\mu( t^{\lambda \mu} + T^{\lambda \mu})\;, \label{10}
\end{equation}
where
\begin{equation}
t^{\lambda \mu}=\kappa\left[4\,\Sigma^{bc\lambda}T_{bc}\,^\mu- g^{\lambda
\mu}\, \Sigma^{abc}T_{abc}\right]\,, \label{11}
\end{equation}
is the gravitational energy-momentum tensor \cite{revmal}. The symmetry of $\Sigma^{a\lambda\nu}$ leads to
\begin{equation}
\partial_\lambda\partial_\nu\left(e\Sigma^{a\lambda\nu}\right)\equiv0\,.\label{12}
\end{equation}
This allows one to define the total energy-momentum vector. It reads
\begin{equation}
P^a = \int_V d^3x \,e\,e^a\,_\mu(t^{0\mu}+ T^{0\mu})\,, \label{14}
\end{equation}
which may be written in the following form
\begin{equation}
P^a =4\kappa\, \int_V d^3x \,\partial_\nu\left(e\,\Sigma^{a0\nu}\right)\,. \label{14.1}
\end{equation}
It is important to point out that the energy-momentum verctor respects the Lorentz symmetry and it is invariant under coordinates transformation.

It is possible to use the Lagrangian density of TEGR above to establish a two points Green function considering the tetrads as the observable fields on space-time. Thus from
\begin{equation}
g_{\mu\nu}=\eta_{\mu\nu}+h_{\mu\nu},
\end{equation}
and expression (\ref{6}) the graviton propagator is \cite{usk}
\begin{equation}
\langle e_{b\lambda}, e_{d\gamma} \rangle=\Delta_{bd\lambda\gamma} = \frac{\eta_{bd}}{\kappa q^{\lambda} q^{\gamma}}.
\end{equation}
Then the Green function reads
\bea
G_0(x,x')=-i\Delta_{bd\lambda\gamma}\,g^{\lambda\gamma}\eta^{bd}.
\eea
Explicitly it is
\begin{equation}
G_0(x,x')= -\frac{i64\pi}{q^{2}}\,,
\end{equation}
with $q=x-x'$, where $x$ and $x'$ are four vectors. With the weak field  approximation the gravitational energy-momentum tensor $t^{\lambda \mu}$ becomes
\begin{eqnarray}
t^{\lambda\mu}(x) &=& \kappa\Bigl[g^{\mu\alpha}\partial^{\gamma}e^{b\lambda}\partial_{\gamma}e_{b\alpha} - g^{\mu\gamma}\partial^{\alpha}e^{b\lambda}\partial_{\gamma}e_{b\alpha} - g^{\mu\alpha}(\partial^{\lambda}e^{b\gamma}\partial_{\gamma}e_{b\alpha} - \partial^{\lambda}e^{b\gamma}\partial_{\alpha}e_{b\gamma})\nonumber\\
        & &-2g^{\lambda\mu}\partial^{\gamma}e^{b\alpha}(\partial_{\gamma}e_{b\alpha}-\partial_{\alpha}e_{b\gamma})\Bigl]\,.
\end{eqnarray}
In order to avoid divergences we adopt the usual procedure to write the energy-momentum tensor at different points in space-time and then taking the proper limit. Hence
\bea
\langle t^{\lambda\mu}(x)\rangle&=& \langle 0|t^{\lambda\mu}(x)|0\rangle,\nonumber\\
&=& \lim_{x^\mu\rightarrow x'^\mu} 4i\kappa\left(-5g^{\lambda\mu}\partial'^{\gamma}\partial_{\gamma} +2g^{\mu\alpha}\partial'^{\lambda}\partial_{\alpha}\right)G_{0}(x-x')\,,\label{em}
\eea
where $\langle e_{c}^{\,\,\,\lambda}(x), e_{b\alpha}(x') \rangle = i\eta_{cb}\,\delta^{\lambda}_{\alpha}\,G_{0}(x-x')$. In this sense it is possible to use the metric of a regular black hole to introduce thermal effects via TFD as explained in the last section. It worths to notice that in the weak field approximation TEGR becomes a usual field which is very different from the metric formulation that cannot dissociate metric from space-time.

\section{Gravitational Casimir effect and Stefan-Boltzmann law at finite temperature for Regular Black Holes} \label{sec.4}

In this section the framework of TFD is used to calculate the mean value of gravitational energy-momentum (\ref{em}) which is obtained in the weak field approximation of TEGR. Thus we have
\bea
\langle t^{\lambda\mu(AB)}(x;\alpha)\rangle=\lim_{x\rightarrow x'} 4i\kappa\left(-5g^{\lambda\mu}\partial'^{\gamma}\partial_{\gamma} +2g^{\mu\alpha}\partial'^{\lambda}\partial_{\alpha}\right)G_{0}^{(AB)}(x-x';\alpha).
\eea
If we use the Casimir prescription,
\bea
{\cal T}^{\lambda\mu (AB)}(x;\alpha)=\langle t^{\lambda\mu(AB)}(x;\alpha)\rangle-\langle t^{\lambda\mu(AB)}(x)\rangle\,,
\eea
then
\bea
{\cal T}^{\lambda\mu (AB)}(x;\alpha)=\lim_{x\rightarrow x'} 4i\kappa\left(-5g^{\lambda\mu}\partial'^{\gamma}\partial_{\gamma} +2g^{\mu\alpha}\partial'^{\lambda}\partial_{\alpha}\right)\overline{G}_{0}^{(AB)}(x-x';\alpha),\label{EM}
\eea
where
\bea
\overline{G}_0^{(AB)}(x-x';\alpha)=G_0^{(AB)}(x-x';\alpha)-G_0^{(AB)}(x-x').
\eea

It is possible to describe a class of regular black holes by the following line element \cite{Neves_Saa}
\begin{equation}
ds^2=-f(r)dt^2+\frac{dr^2}{f(r)}+r^2 \left( d\theta^2+\sin^2\theta d\phi^2 \right),
\label{Metric}
\end{equation}
with
\begin{equation}
f(r)=1-\frac{2\, M_0}{r \left[1+\left(\frac{r_0}{r}\right)^{q} \right]^{\frac{p}{q}}}\,,
\end{equation}
where $M_0$ is the the mass of the regular black hole, in fact it coincides with ADM mass in the limit $r\rightarrow \infty$. The parameter $r_0$ can be seen as a fundamental length of the regular black hole. Such a line element reproduces known solutions for a suitable choice of the parameters $p$ and $q$ which has to be integers. For instance the Bardeen solution is achieved for $p=3$ and $q=2$, while Hayward solution for $p=q=3$. Thus it should be noted that the metric in eq. (30) represents a class of solutions. Such regular black holes arose in order to deal with an open problem in gravitation, the existence of singularities. In fact the Bardeen solution was the first class of regular black holes which can be understood as a magnetic monopole coupled to Einstein equation\cite{bardeenmonopole}. It is important to point out that although the metric in eq. (30) has no singularity at $r=0$, it does have an event horizon given by the solution of $f(R_H)=0$. On the other hand the stability of such solutions need to be investigated.

\subsection{Gravitational Stefan-Boltzmann Law }
In order to analyze the gravitational Stefan-Boltzmann law we have to choose
$\alpha=(\beta,0,0,0)$ in the TFD formalism. Then the Bogoliubov transformation  is
\bea
v^2(\beta)=\sum_{j_0=1}^\infty e^{-\beta k^0 j_0}.
\eea
Hence we have to use the following Green function
\bea
\overline{G}_0^{(11)}(x-x';\beta)&=&2\sum_{j_0=1}^\infty G_0^{(11)}\left(x- x'-i\beta j_0 n_0\right),\label{1GF}
\eea
where $n_0=(1,0,0,0)$. Thus from (\ref{EM}) we can calculate the energy with $(AB)=(11)$ which is the physical component in the matrix obtained in TFD. It reads
\begin{eqnarray}
 {\cal T}^{00(11)}(x;\alpha)&=&\epsilon(r, T) =\lim_{x\rightarrow x'}\sum_{j_0=1}^{\infty} 4\kappa i\Biggl\{-3\left[ 1+ \frac{4\, M_0}{r \left[1+\left(\frac{r_0}{r}\right)^{q} \right]^{\frac{p}{q}}} \right]\partial'_{0}\partial_{0} +5\partial'_{1}\partial_{1}\nonumber\\
&+&\frac{5}{r^2}\left[ 1 + \frac{2\, M_0}{r \left[1+\left(\frac{r_0}{r}\right)^{q} \right]^{\frac{p}{q}}}\right]\left( \partial'_{2}\partial_{2}+\dfrac{1}{\sin ^{2}{\theta}} \partial'_{3}\partial_{3} \right) \Biggr\}\, G_0^{(11)}\left(x- x'-i\beta j_0 n_0\right)\,,
\end{eqnarray}
Once the Riemann zeta function is defined by
$\zeta(4)=\sum_{j_0=1}^\infty\frac{1}{j_0^4}=\frac{\pi^4}{90},\label{zetaf}$
then the gravitational Stefan-Boltzmann energy is
\begin{equation}
 \epsilon(r, T)= \dfrac{32 \pi^{4}}{15}\left\{ 1+ \frac{6\, M_0}{r \left[1+\left(\frac{r_0}{r}\right)^{q} \right]^{\frac{p}{q}}} \right\}T^{4}\,,
\end{equation}
here we have to notice that for the vanishing of the physical parameter of the regular black hole $M_0$ the energy does not goes to zero. Therefore we need to regularize such an expression by requiring
$E(r,T)= \epsilon(r, T)-\dfrac{32\pi^{4}T^4}{15}$. That leads to
\begin{equation}
E(r,T)= \left\{ \frac{192\pi^{4}\, M_0}{15\,r \left[1+\left(\frac{r_0}{r}\right)^{q} \right]^{\frac{p}{q}}} \right\}T^{4}\,,
\end{equation}
which is the regularized energy.

The first law of Thermodynamics states
\begin{equation}
    E + P = T\left(\dfrac{\partial P}{\partial T} \right)_{V}\,,
\end{equation}
then if we use the regularized energy above such first order differential equation has the solution  $P=\frac{E}{3}$. This is very interesting to note that is equal to the photon state equation. In terms of the temperature the pressure is explicitly
\begin{equation}
    P(r,T) = \left\{ \frac{192\pi^{4}\, M_0}{45\,r \left[1+\left(\frac{r_0}{r}\right)^{q} \right]^{\frac{p}{q}}} \right\} T^{4}.
\end{equation}

In order to calculate the gravitational entropy we recall the definition $P=-\dfrac{\partial F}{\partial V}$, where $F$ is the free energy, and  $S=-\dfrac{\partial F}{\partial T}$. Therefore from
$\left(\dfrac{\partial P}{\partial T} \right)_{V}=\left(\dfrac{\partial S}{\partial V} \right)_{T}$ the entropy is
\begin{eqnarray}
    S&=&\dfrac{3072 M_0\pi^{5}}{45}T^{3}\int_0^R \left\{ \frac{r}{\left[1+\left(\frac{r_0}{r}\right)^{q} \right]^{\frac{p}{q}}} \right\}  dr \,,
\end{eqnarray}
which becomes
\begin{equation}
S=\dfrac{3072 M_0\pi^{5}R^2}{90}\,{}_2F_1\left(-\frac{2}{q},\frac{p}{q};1-\frac{2}{q}; \left(\frac{r_0}{R}\right)^{q}\right)\, T^{3} \,,
\end{equation}
where ${}_2F_1 $ is the hypergeometric function. The integration hyper-surface is a sphere of radius $R$. Thus the gravitational entropy exists on an arbitrary portion of space which is a different approach from the usual black hole Thermodynamics. Usually a black hole has a fixed entropy written in terms of its event horizon area. Here the following principles are assumed: i) the entropy is a function of macroscopic parameters. If the ``no hair'' theorems are valid\cite{nohair}, then the entropy is a function of mass, angular momentum and charge. ii) The Penrose process leads to an arbitrary manipulation of these parameters provided that the event horizon area remains unchanged. That holds for regular black holes\cite{penroserbh}. It implies that the entropy has to be a function of area. It should be noted the metric of the regular black hole tends to Schwarzschild metric for a position far from the event horizon. Therefore the entropy tends to be proportional to the event horizon area, hence it is reasonable to admit that $S_H=A_H/4$ for a regular black hole, where $A_H=4\pi R_H^2$, with $R_H$ being  the solution of $$\frac{2\, M_0}{R_H \left[1+\left(\frac{r_0}{R_H}\right)^{q} \right]^{\frac{p}{q}}} =1\,.$$ Therefore the temperature of the event horizon of a class of regular black holes defined by (\ref{Metric}) is

\begin{equation}
T_H =\left[\frac{90}{3072 M_0\pi^{4}\,{}_2F_1\left(-\frac{2}{q},\frac{p}{q};1-\frac{2}{q}; \left(\frac{r_0}{R_H}\right)^{q}\right)}\right]^{1/3}\,,
\end{equation}
which is a unique prediction of TFD applied to TEGR.

\subsection{Gravitational Casimir Effect}
If the Casimir effect description is desired then the choice
 $\alpha=(0,i2d,0,0)$ has to be made which leads to the following Bogoliubov transformation
\bea
v^2(d)=\sum_{l_1=1}^\infty e^{-i2dk^1l_1}.
\eea
If the Green function is given by
\bea
\overline{G}_0^{(11)}(x-x';d)&=&2\sum_{l_1=1}^\infty G_0^{(11)}\left(x-x'-2dl_1r\right),\label{2GF}
\eea
then
\begin{eqnarray}
{\cal T}^{00 (11)}(d, r) &=& \epsilon_{c}(d, r)=\lim_{x\rightarrow x'}\sum_{l_1=1}^{\infty} 4\kappa i\Biggl\{-3\left[ 1+ \frac{4\, M_0}{r \left[1+\left(\frac{r_0}{r}\right)^{q} \right]^{\frac{p}{q}}} \right]\partial'_{0}\partial_{0} +5\partial'_{1}\partial_{1}\nonumber\\
&+&\frac{5}{r^2}\left[ 1 + \frac{2\, M_0}{r \left[1+\left(\frac{r_0}{r}\right)^{q} \right]^{\frac{p}{q}}}\right]\left( \partial'_{2}\partial_{2}+\dfrac{1}{\sin ^{2}{\theta}} \partial'_{3}\partial_{3} \right) \Biggr\}\, G_0^{(11)} \left(x- x'-2dl_1 n_1\right)\,,
\end{eqnarray}
where $n_1=(0,1,0,0)$. Hence the energy associated to the gravitational Casimir effect for regular black holes is
\begin{equation}
    \epsilon_{c}(d, r) = \sum_{l_1} -\dfrac{4}{d^{4}l_1^{4}}\left\{1-\frac{2\, M_0}{r \left[1+\left(\frac{r_0}{r}\right)^{q} \right]^{\frac{p}{q}}} +\left[\frac{2\, M_0}{r \left[1+\left(\frac{r_0}{r}\right)^{q} \right]^{\frac{p}{q}}}\right]\left(\dfrac{5dl_1}{r}\right)\right\}\,,
\end{equation}
which for the approximation $d<<r$ becomes
\begin{eqnarray}
    \epsilon_{c}(d, r) &=& -\dfrac{2\pi^{4}}{45d^{4}}\left\{1-\frac{2\, M_0}{r \left[1+\left(\frac{r_0}{r}\right)^{q} \right]^{\frac{p}{q}}}\right\}\,.
\end{eqnarray}
It should be noted that the vacuum contribution is negative and has a dependency of $d^{-4}$. In order to take into account only the regular black hole contribution, a regularization procedure is necessary. Thus subtracting the vacuum energy we have
$$E_{c}(d, r) =\left\{\frac{4 \pi^{4}\, M_0}{45d^{4} \,r \left[1+\left(\frac{r_0}{r}\right)^{q} \right]^{\frac{p}{q}}}\right\} \,,$$ where $E_{c}(d, r)$ is the regularized Casimir energy. It is interesting to note that on the event horizon it is exactly minus the vacuum energy.

Similarly  the Casimir pressure is
\begin{eqnarray}
{\cal T}^{33 (11)}(d, r) &=& \rho_{c}(d, r) =\lim_{x\rightarrow x'}\sum_{l_1=1}^{\infty} 4\kappa i\left[1-\frac{2\, M_0}{r \left[1+\left(\frac{r_0}{r}\right)^{q} \right]^{\frac{p}{q}}} \right]\Bigg\{ 5\left[ 1+ \frac{2\, M_0}{r \left[1+\left(\frac{r_0}{r}\right)^{q} \right]^{\frac{p}{q}}} \right]\partial'_{0}\partial_{0}\nonumber\\
&-&3\left[ 1-\frac{2\, M_0}{r \left[1+\left(\frac{r_0}{r}\right)^{q} \right]^{\frac{p}{q}}} \right]\partial'_{1}\partial_{1}-\dfrac{5}{r^{2}} \left(\partial'_{2}\partial_{2}+\dfrac{1}{\sin ^{2}{\theta}}\partial'_{3}\partial_{3}  \right) \Bigg\}\, G_0^{(11)} \left(x- x'-2dl_1 n_1\right)\,.\nonumber\\
\end{eqnarray}
This yields
\begin{eqnarray}
    \rho_{c}(d, r )=-\dfrac{4}{d^{4}} \sum_{l_1} \dfrac{1}{l_1^{4}} \left\{ 3-\frac{18\, M_0}{r \left[1+\left(\frac{r_0}{r}\right)^{q} \right]^{\frac{p}{q}}} -\frac{6dl_1\, M_0}{r^2 \left[1+\left(\frac{r_0}{r}\right)^{q} \right]^{\frac{p}{q}}} \right\}\,,
\end{eqnarray}
which, after the limit $\dfrac{d}{r} \ll 1$ is taken, reads
\begin{equation}
    \rho_{c}(d, r) =-\dfrac{2\pi^{4}}{15d^{4}}\left\{1-\frac{6\, M_0}{r \left[1+\left(\frac{r_0}{r}\right)^{q} \right]^{\frac{p}{q}}}\right\}\,.
\end{equation}
Again in order to consider the non-vacuum contribution a regularized pressure is necessary. Thus the regularized gravitational Casimir pressure is
$$P_{c}(d, r) =\left\{\frac{12 \pi^{4}\, M_0}{15d^{4} \,r \left[1+\left(\frac{r_0}{r}\right)^{q} \right]^{\frac{p}{q}}}\right\}\,.$$ It should be noted that both the regularized Casimir energy and pressure are very small due to the weak field approximation. We would like to point out that in this formalism the vacuum itself has some gravitational features which explain why a regularization is necessary.

\subsection{Gravitational Casimir Effect at Finite Temperature}
The choice $\alpha=(\beta, i2d, 0, 0)$ is suitable to describe the Casimir effect at finite temperature. As a consequence following the TFD prescription the Bogoliubov transformation is given by
\bea
v^2(k^0,k^3;\beta,d)&=&v^2(k^0;\beta)+v^2(k^1;d)+2v^2(k^0;\beta)v^2(k^1;d),\nonumber\\
&=&\sum_{j_0=1}^\infty e^{-\beta k^0j_0}+\sum_{l_1=1}^\infty e^{-i2dk^1l_1}+2\sum_{j_0,l_1=1}^\infty e^{-\beta k^0j_0-i2dk^1l_1}\,,\label{eq51}
\eea
where the first term takes into account temperature effects, the second term stands for the Casimir effect only and the last term the interaction between both. The Green function is then
\bea
\overline{G}_0^{(11)}(x-x';\beta,d)&=&4\sum_{j_0,l_1=1}^\infty G_0^{(11)}\left(x-x'-i\beta j_0n-2dl_1r\right)\,.\label{3GF}
\eea
As before the gravitational Casimir energy is obtained from expression (\ref{EM}) which reads
\bea
\epsilon_{c}(\beta,d)&=&\lim_{x\rightarrow x'}\sum_{j_0,l_1=1}^\infty 4\kappa i\Biggl\{-3\left[ 1+ \frac{4\, M_0}{r \left[1+\left(\frac{r_0}{r}\right)^{q} \right]^{\frac{p}{q}}} \right]\partial'_{0}\partial_{0} +5\partial'_{1}\partial_{1}\nonumber\\
&+&\frac{5}{r^2}\left[ 1 + \frac{2\, M_0}{r \left[1+\left(\frac{r_0}{r}\right)^{q} \right]^{\frac{p}{q}}}\right]\left( \partial'_{2}\partial_{2}+\dfrac{1}{\sin ^{2}{\theta}} \partial'_{3}\partial_{3} \right) \Biggr\}\, G_0^{(11)}\left(x-x'-i\beta j_0n-2dl_1r\right)\,,\nonumber\\
\eea
where $\epsilon_{c}(\beta, d)={\cal T}^{00 (11)}(\beta;d)$. It worths to obtain the regularized Casimir energy at finite temperature which is achieved by subtracting the vacuum energy from $\epsilon_{c}(\beta,d)$, explicitly it is
\begin{eqnarray}
    E_{c}(\beta, d) &=& -64\sum_{j_0,l_1=1}^\infty \left[\frac{1}{4d^{2} l_1 ^{2}\left(1+ \frac{2\, M_0}{r \left[1+\left(\frac{r_0}{r}\right)^{q} \right]^{\frac{p}{q}}} \right)+ j_0^{2}\left( 1-\frac{2\, M_0}{r \left[1+\left(\frac{r_0}{r}\right)^{q} \right]^{\frac{p}{q}}}\right)\beta^{2}} \right]^{3} \Bigg\{4d^{2}l_1^{2}\Bigg[1\nonumber\\
    &+&12\left(\frac{\, M_0}{r \left[1+\left(\frac{r_0}{r}\right)^{q} \right]^{\frac{p}{q}}}\right)^{2}+\left[\frac{2\, M_0}{r \left[1+\left(\frac{r_0}{r}\right)^{q} \right]^{\frac{p}{q}}}\right]\left(1+\frac{5d l_1}{r} \right)\Bigg] \left( 1+ \frac{2\, M_0}{r \left[1+\left(\frac{r_0}{r}\right)^{q} \right]^{\frac{p}{q}}}\right) \nonumber\\
    & &-j_0^{2}\Bigg[3 -72\left(\frac{\, M_0}{r \left[1+\left(\frac{r_0}{r}\right)^{q} \right]^{\frac{p}{q}}}\right)^{3}+30dl_1 \left(\frac{\, M_0}{r^2 \left[1+\left(\frac{r_0}{r}\right)^{q} \right]^{\frac{p}{q}}}\right)\nonumber\\&+&4\left(\frac{ M_0}{r \left[1+\left(\frac{r_0}{r}\right)^{q} \right]^{\frac{p}{q}}}\right)^{2}\left(6+\frac{5dl_1}{r}\right) \Bigg]\beta^{2}\Bigg\}+64\sum_{j_0,l_1=1}^\infty\frac{4d^{2} l_1 ^{2}-3j_0^{2}\beta^{2}}{\left(4d^{2} l_1 ^{2}+ j_0^{2}\beta^{2}\right)^3}\,,
\end{eqnarray}
where $E_{c}(\beta, d)$ is the regularized expression. Similarly the gravitational Casimir pressure, $\rho_{c}(\beta,d)$, at finite temperature is given by
\begin{eqnarray}
\rho_{c}(\beta,d) &=&\lim_{x\rightarrow x'} \sum_{j_0,l_1=1}^\infty 4\kappa i\left[1-\frac{2\, M_0}{r \left[1+\left(\frac{r_0}{r}\right)^{q} \right]^{\frac{p}{q}}} \right]\Bigg\{ 5\left[ 1+ \frac{2\, M_0}{r \left[1+\left(\frac{r_0}{r}\right)^{q} \right]^{\frac{p}{q}}} \right]\partial'_{0}\partial_{0}\nonumber\\
&-&3\left[ 1-\frac{2\, M_0}{r \left[1+\left(\frac{r_0}{r}\right)^{q} \right]^{\frac{p}{q}}} \right]\partial'_{1}\partial_{1}-\dfrac{5}{r^{2}} \left(\partial'_{2}\partial_{2}+\dfrac{1}{\sin ^{2}{\theta}}\partial'_{3}\partial_{3}  \right) \Bigg\}\, G_0^{(11)}\left(x-x'-i\beta j_0n-2dl_1r\right)\,,\nonumber\\
\end{eqnarray}
where $\rho_{c}(\beta, d)={\cal T}^{11 (11)}(\beta;d) $. As the regularized energy, the regularized Casimir pressure is
\begin{eqnarray}
    P_{c}(\beta,d) &=& -64\sum_{j_0,l_1=1}^\infty \left[\frac{1}{4d^{2} l_1 ^{2}\left(1+ \frac{2\, M_0}{r \left[1+\left(\frac{r_0}{r}\right)^{q} \right]^{\frac{p}{q}}} \right)+ j_0^{2}\left( 1-\frac{2\, M_0}{r \left[1+\left(\frac{r_0}{r}\right)^{q} \right]^{\frac{p}{q}}}\right)\beta^{2}} \right]^{3}\Bigg\{4d^{2} l_1 ^{2}\Bigg[3\nonumber\\
 &+&8\left(\frac{M_0}{r \left[1+\left(\frac{r_0}{r}\right)^{q}\right]}\right)^{2} -6d l_1 \left(\frac{M_0}{r^2 \left[1+\left(\frac{r_0}{r}\right)^{q} \right]}\right)\left(1-\left(\frac{2\, M_0}{r \left[1+\left(\frac{r_0}{r}\right)^{q} \right]}\right)\right)\Bigg] \Bigg[1\nonumber\\
 &-&\left(\frac{2M_0}{r \left[1+\left(\frac{r_0}{r}\right)^{q}\right]}\right)^{2} \Bigg]+j_0 ^{2}\Bigg[ 24 \left(\frac{M}{r}\right)^2+6dl_1 \left(\frac{M}{r^2}\right)\left(3+ \left(\frac{2M}{r}\right)\right) -1 \Bigg] \Bigg[ 1\nonumber\\
 &-&\left(\frac{2M}{r}\right) \Bigg]^{2} \beta^{2}\Bigg\}+ 64\sum_{j_0,l_1=1}^\infty\frac{4d^{2} l_1 ^{2}-j_0^{2}\beta^{2}}{\left(4d^{2} l_1 ^{2}+ j_0^{2}\beta^{2}\right)^3}\,,
\end{eqnarray}
where $P_{c}(\beta,d) = \rho_{c}(\beta,d) + 64\sum_{j_0,l_1=1}^\infty\frac{4d^{2} l_1 ^{2}-j_0^{2}\beta^{2}}{\left(4d^{2} l_1 ^{2}+ j_0^{2}\beta^{2}\right)^3}\,. $ The regularized expressions take into account only the contributions of the regular black holes. They are small corrections to the vacuum quantities which have the known limits for $\beta\rightarrow\infty$. It should be noted that the gravitational Casimir effect is very controversial idea due to the energy problem in general relativity. As a matter of fact the lack of a gravitational energy-momentum tensor in this approach prevents one from analyzing such effect. On the other hand in the framework of TEGR the gravitational Casimir effect can be explored.

\section{Conclusion} \label{sec.5}

The regular black holes are studied at finite temperature. The temperature effects are introduced using the TFD formalism. TFD is a tool that allows to analyze temperature effects in addition to the time dependence. Using the Teleparalelism Equivalent to General Relativity the gravitational thermodynamics to the regular black holes is investigated. This gravitational theory has a well defined energy-momentum tensor that allows to calculated the gravitational Stefan-Boltzmann law and Casimir effect associated to the regular black holes. A regularized gravitational Stefan-Boltzmann law for regular black hole is obtained. Using the first law of thermodynamics the gravitational pressure and the gravitational entropy are determined. The relation between gravitational energy and pressure is equal the relation that describes the photon. The gravitational entropy obtained here exists on an arbitrary portion of space. Then it is a different approach from the usual black hole thermodynamics, since the usual black hole has a fixed entropy given in terms of its event horizon area. In addition, the temperature of the event horizon for regular black holes has been calculated. Furthermore, the gravitational Casimir energy and Casimir pressure at zero and finite temperature for this class of regular black holes are determined. It is interesting to note that such results can be experimentally verified, once confirmed, it suggests that the torsion tensor is the true quantity responsible by gravitation instead of curvature as the mainstream approach for the gravitational field.

\section*{Acknowledgments}

This work by A. F. S. is supported by CNPq projects 308611/2017-9 and 430194/2018-8. A. E. S. also thanks CNPq (a federal Brazilian Government Agency) for partial financial support.

\end{document}